\documentclass[%
 reprint,
 amsmath,amssymb,
 aps,
]{revtex4-1}

\usepackage{color}

\usepackage{ragged2e}
\usepackage{hyperref}
\usepackage{graphicx}
\usepackage{dcolumn}
\usepackage{bm}
\usepackage{color}

\begin{document}
\bibliographystyle{naturemag}

\preprint{APS/123-QED}

\title{Network models in neuroscience}

\author{Danielle S. Bassett$^{1,2,3,4,7}$}
\author{Perry Zurn$^{5}$}
\author{Joshua I. Gold$^{6}$}
\affiliation{
 $^1$Department of Bioengineering, School of Engineering and Applied Sciences, University of Pennsylvania, Philadelphia, PA, 19104
}
\affiliation{
 $^2$Department of Physics \& Astronomy, College of Arts and Sciences, University of Pennsylvania, Philadelphia, PA, 19104
}
\affiliation{
$^3$Department of Electrical \& Systems Engineering, School of Engineering and Applied Sciences, University of Pennsylvania, Philadelphia, PA, 19104
}
\affiliation{
 $^4$Department of Neurology, Perelman School of Medicine, University of Pennsylvania, Philadelphia, PA, 19104
}
\affiliation{
 $^5$Department of Philosophy, American University, Washington, DC, 20016
}
\affiliation{
 $^6$Department of Neuroscience, Perelman School of Medicine, University of Pennsylvania, Philadelphia, PA, 19104
}
\affiliation{
$^7$To whom correspondence should be addressed: dsb@seas.upenn.edu
}

\begin{abstract}From interacting cellular components to networks of neurons and neural systems, interconnected units comprise a fundamental organizing principle of the nervous system. Understanding how their patterns of connections and interactions give rise to the many functions of the nervous system is a primary goal of neuroscience. Recently, this pursuit has begun to benefit from the development of new mathematical tools that can relate a system's architecture to its dynamics and function. These tools, which are known collectively as network science, have been used with increasing success to build models of neural systems across spatial scales and species. Here we discuss the nature of network models in neuroscience. We begin with a review of model theory from a philosophical perspective to inform our view of networks as models of complex systems in general, and of the brain in particular. We then summarize the types of models that are frequently studied in network neuroscience along three primary dimensions: from data representations to first-principles theory, from biophysical realism to functional phenomenology, and from elementary descriptions to coarse-grained approximations. We then consider ways to validate these models, focusing on approaches that perturb a system to probe its function. We close with a description of important frontiers in the construction of network models and their relevance for understanding increasingly complex functions of neural systems. \end{abstract}

\maketitle

\section*{Introduction}
 
\noindent The brain is composed of intricate networks that operate at many different levels of organization. For example, at small spatial scales gene regulatory networks direct neuronal cell fate, and both chemical and electrical synapses define the accessible routes of information transmission between neurons \cite{francis2003bridging}. At intermediate spatial scales, laminar architecture in cortex is accompanied by stereotyped inter-laminar connectivity thought to support ensemble dynamics and resultant computations \cite{sherman2016neural}. At even larger spatial scales, the anatomical locations of inter-areal projections display a precise spatial arrangement associated with a diverse repertoire of functional processes \cite{betzel2018specificity,betzel2017diversity}. 

Although networks are fundamental to brain structure, the complexity of these networks poses challenges to understanding their function. Unlike a sphere, which we can quickly guess to have the capacity to be rolled or thrown, a network -- with its tangle of wires -- defies any simple conspectus. Thus, even though more than a century has passed since Camillo Golgi, Santiago Ramon y Cajal, and other neuroanatomists introduced to the world the intricate beauty of the networks of neurons that comprise our nervous system, our understanding of how those networks give rise to perception, learning, memory, cognition, action, and other aspects of brain function remains incomplete.

One fruitful set of approaches for understanding relationships between the brain's networked architecture and its many functions that has emerged in recent years is network science. This discipline deals with the study of systems whose structure, function, or dynamics depend upon the pattern of interconnections between units \cite{albert2002statistical}. Network science is inherently interdisciplinary, drawing on and integrating among recent advances in mathematics, physics, computer science, and engineering \cite{newman2010networks,newman2011complex}. Although early work in the field was largely devoted to the study of social systems, efforts over the last decade have focused increasingly on the study of neural systems across spatial scales, temporal scales, and species \cite{bullmore2009complex,fornito2016brain,heuvel2016comparative}. These newer efforts, collectively referred to as \emph{network neuroscience}, model neural systems as networks to distill the dependence of brain function and dysfunction on interconnection architecture \cite{bassett2017network}.

In this chapter, we review recent work in network neuroscience that has been applied to understanding the brain, emphasizing the diversity of approaches that now fall under this general framework \cite{bassett2018nature}. Because network neuroscience is fundamentally a modelling endeavor, we begin with a broad perspective on model theory from philosophy. We then consider how networks are models, before turning to a discussion of the types of network models that are commonly used in neuroscience. Motivated by approaches to validate network models via prediction, we discuss the importance of perturbation-based techniques for understanding network function. We close by outlining important directions for future work to build, use, and validate network models in neuroscience.\\
 
\section*{Model Theory - A philosophical perspective}

The term ``model'' can bring quite different pictures to the mind's eye. A small, inexact replica of a 1910 Schacht Roadster. A miniature cityscape coarsely true to the form of Cambridge, England. Rodin's ``The Thinker''. A contemporary reworking of a piece from Greek mythology \cite{eugenides2002middlesex}. A person one wishes to emulate, or an ideal one hopes to become. IBM's Watson. A cerebral organoid (miniature organ \emph{in vitro} \cite{huang2017tranylcypromine}) or a human blinking eye-on-a-chip \cite{chan2015vitro}. A set of interdependent partial differential equations producing dynamics reminiscent of a real-world system. In mentally traversing these diverse examples, one immediately realizes that the space of model types is exceedingly large, and one wonders whether it is even possible to define what a model actually is, or what it is not.

The question of how to define the term ``model'' is the focus of a branch of philosophy known as model theory, which aims to identify the essential elements that make models what they are and to disambiguate the characteristics that distinguish different types of models from one another \cite{gelfert2016how}. At its most basic level, a model is a representation of one or more aspects of the world. It aims to better understand what something is by measuring and imaging what something does. As such, models inherit a basic philosophical conundrum \cite{papineau1987reality}: what precisely is their relationship to the target systems they model? And from whence do they derive their truth value? Must they simply evidence functional coherence and have pragmatic purchase, or must they also meaningfully correspond to what they represent, and, if so, how is that meaningfulness determined?

In science, at least four types of models have been recognized that each provide different answers to these questions. These types are: scalar, idealized, analogical, and phenomenological \cite{hartmann2012scientific,frigg2012models}. Scalar models, much like the Roadster replica, either magnify or reduce their target systems. Idealized models abstract and isolate a limited set of features from their target systems. Analogical models highlight relevant similarities between two target systems, whether those similarities are shared properties or comparable structures. Finally, phenomenological models represent only the observable elements of their target systems, without postulating any theoretical explanation as to why those elements are what they are. 

Recent work in model theory explores the intriguing possibility that all of these forms of scientific models are heuristic devices not unlike literary fictions \cite{toon2012models,frigg2010mimesis,suarez2009fictions}. The models-as-fictions theory reconceptualizes models as fictional entities that aim to narrativize certain features of a target system \cite{barbrousse2009fictions,frigg2010models,frigg2016fiction,godfrey2009models,toon2012tools,frigg2010fiction,frigg2010fictions,garcia2010fictional,frigg2017scientific}. According to this framework, models -- much like fictions -- may imaginatively isolate and abstract or distort and exaggerate certain features of the world in such a way as to facilitate epistemic access \cite{elgin2010telling}. They may creatively instantiate either analogical or phenomenological substructures of their target systems in order to crystallize insight. Scientific models are therefore subject to evaluation at the level of both artistry (clarity, elegance, originality) and function. Moreover, just as the literary tradition provides new fiction with meaningful constraints in advance, so the scientific community provides the parameters within which new models are developed and applied. Scientific models are, in this sense, accountable to the scientific communities that use them in the exploration of target systems that are already of particular value and interest \cite{almeder2007pragmatism}. Finally, different sorts of models can be used together to build a multi-scalar narrative architecture, modeling complementary features of a target system beside one another. Whether used in isolation or in conjunction, scientific models illuminate the overarching structure of a target system precisely through the practice and provocation of creative imagination.

\section*{Networks as models}
\noindent The incipient challenge in modeling biological systems is to identify the most meaningful characteristics of the system that are distilled into a sensible representation \cite{bellomo2015on}. That is, biological models are inherently idealized models of complex systems, and their construction requires identifying first the form and degree of abstraction to use. This process requires a set of value judgments (``What are the most meaningful characteristics?'') and a commitment to epistemic cleanliness (``What details of biology can we defensibly ignore?''). These principles of valuation and purposeful ignorance are manifest even when exercising simple visual depictions, which arguably comprise the most impoverished of modeling approaches \cite{tufte2001visual}. One similarly faces choices of what to depict and what not to depict when building any simple mathematical representation of the system. For example, when building a differential equation to represent a system, one must choose which processes to encapsulate and not encapsulate in a variable. 

The fundamental assumption of network neuroscience is that idealized models of the brain should be constructed using analogical principles that focus on the networked architecture of the nervous system. As Cajal saw under his microscope, the nervous system is composed of individual neurons that are interconnected in complex ways. Accordingly, the earliest network models were idealized versions of this network structure, with nodes representing neurons and edges representing the connections between them. More recently, network models have been developed within and across multiple spatial and temporal scales, at not just the level of interconnected neurons but also networks of subcellular components, multi-cellular systems, or both. As detailed below, these models also can be phenomenological, based on measured elements of the nervous system, or more theory driven. Despite this diversity, these models retain key features that can be understood in terms of their basic idealized and analogical structure: an architecture built using interconnected units.

Such an architecture is typically encoded in a graph: an object composed of nodes, which represent units of the system, and edges, which represent interactions or links between those units \cite{bollobas1979graph,bollobas1985random}. Studies of graphs can be neatly separated into two categories: those that consider artificial graphs with arbitrary wiring principles \cite{harary1969graph}, and those that consider graphs that reflect the architecture of a real system \cite{cohen2010complex}. In both cases, one seeks to describe the mathematical properties of the graph with the goal of understanding the function of the system. The patterns of which units can and cannot (or do and do not) interact with one another can allow one to deduce where information might be relatively more densely or relatively more sparsely located, where vulnerability might exist to injury or perturbation, and where circumscribed instances of collective dynamics might emerge \cite{gomez2007paths,cisneros2002information,simonsen2008transient,albert2000error}.

\begin{figure*}[ht]
\centering
\includegraphics[width=1\linewidth]{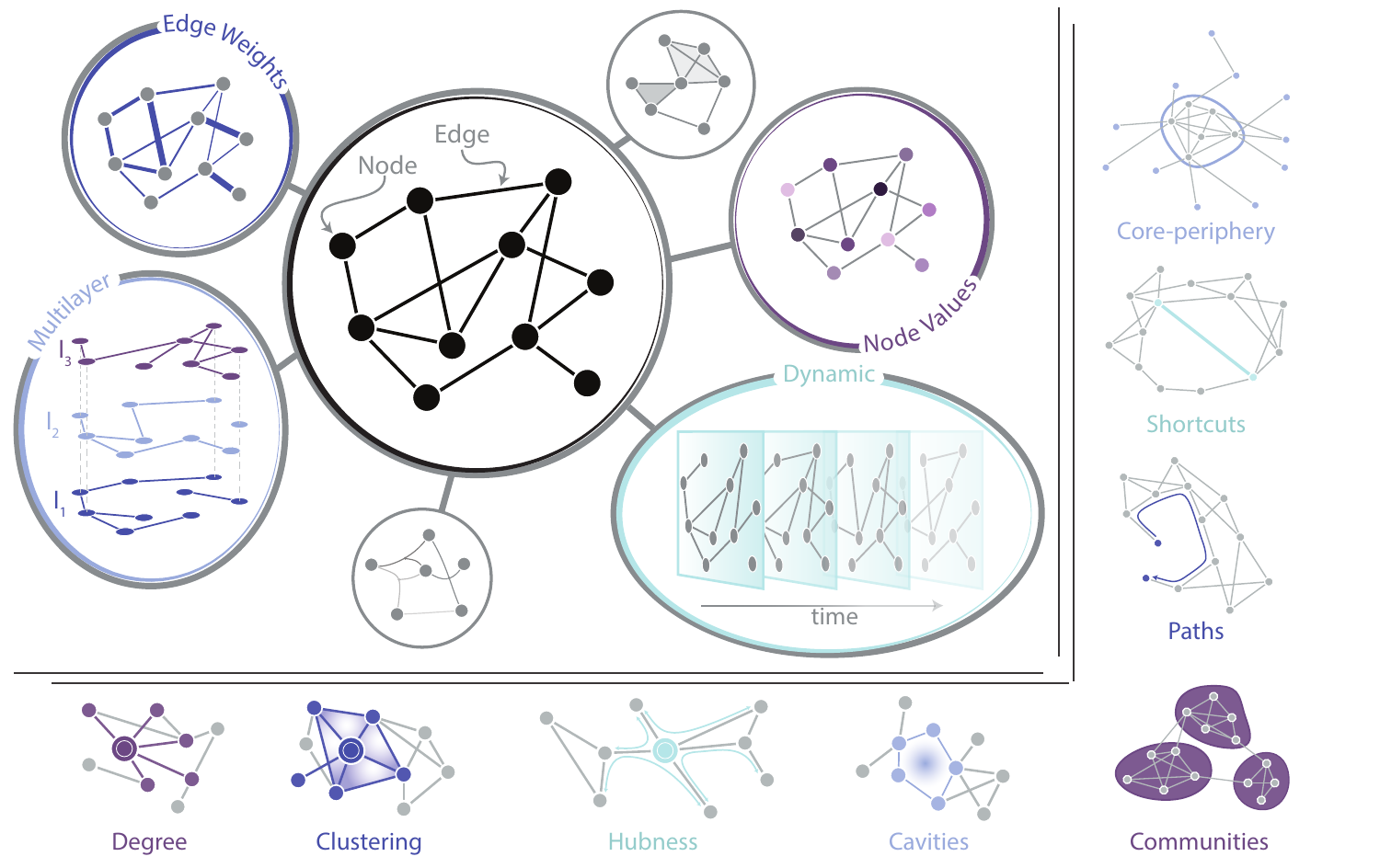}
\caption{\textbf{Schematic of network models.} \emph{(Top Left Corner)} The simplest network model for neural systems is one that represents the pattern of connections (edges) between neural units (nodes). More sophisticated network models can be constructed by adding edge weights and node values, or explicit functional forms for their dynamics. Multilayer networks can be used to represent a set of interconnected networks, and dynamic networks can be used to understand the reconfiguration of network systems over time. \emph{(Bottom Right Corner)} Common measures of interest include: degree, which is the number of edges emanating from a node; clustering, which is related to the prevalence of triangles; cavities, which describe the absence of edges; hubness, which is related to a node's influence; paths, which determine the potential for information transmission; communities, or local groups of densely interconnected nodes; shortcuts, which are one possible marker of global efficiency of information transmission; and core-periphery structure, which facilitates local integration of information gathered from or sent to more sparsely connected areas. Figure adapted with permission from \cite{bassett2018nature}.}
\label{fig1}
\end{figure*}

In simple graphs, all units are represented by identical nodes, and all edges are represented as either existing or not existing (Fig.~\ref{fig1}). These representations can be encoded using a binary weighting scheme. Furthermore, interactions are assumed to be bidirectional: if an edge exists between node $i$ and node $j$, then an edge also exists between node $j$ and node $i$. The very first formal network models of neural systems employed such binary, undirected graphs \cite{felleman1991distributed,young1994analysis,scannell1995analysis,achard2006resilient,sporns2005human,kaiser2006nonoptimal}. Nonetheless, it is relatively straightforward to adapt this encoding to a continuous weighting scheme, as well as to specify distinct weights for the edge from node $i$ to node $j$, and for the edge from node $j$ to node $i$. With the continued refinement of empirical measurement techniques, the inclusion of edge weights in network models has become increasingly prevalent, providing richer insights into system function and dynamics \cite{bassett2016small,rubinov2011weight,betzel2018specificity,markov2013role,oh2014mesoscale}.

Network models are simple constructs. They can be used effectively to study social, biological, technological, and physical systems \cite{newman2010networks}. Yet this flexibility is a marked reminder that the intuitions that one gains from a network model depend strongly on what the nodes and edges are chosen to represent. A structural motif in a network of humans interlinked by friendships can mean something quite different than the same structural motif in a network of neurons interlinked by synapses. Thus, in any endeavor that translates a complex system into a network model, it is critical to specify exactly what the nodes and edges (or more complicated model components) represent, and to ensure that interpretations are drawn in accordance with those choices \cite{butts2009revisiting}.

\section*{Types of Network Models in Neuroscience}

\begin{figure*}[ht]
\centering
\includegraphics[width=1\linewidth]{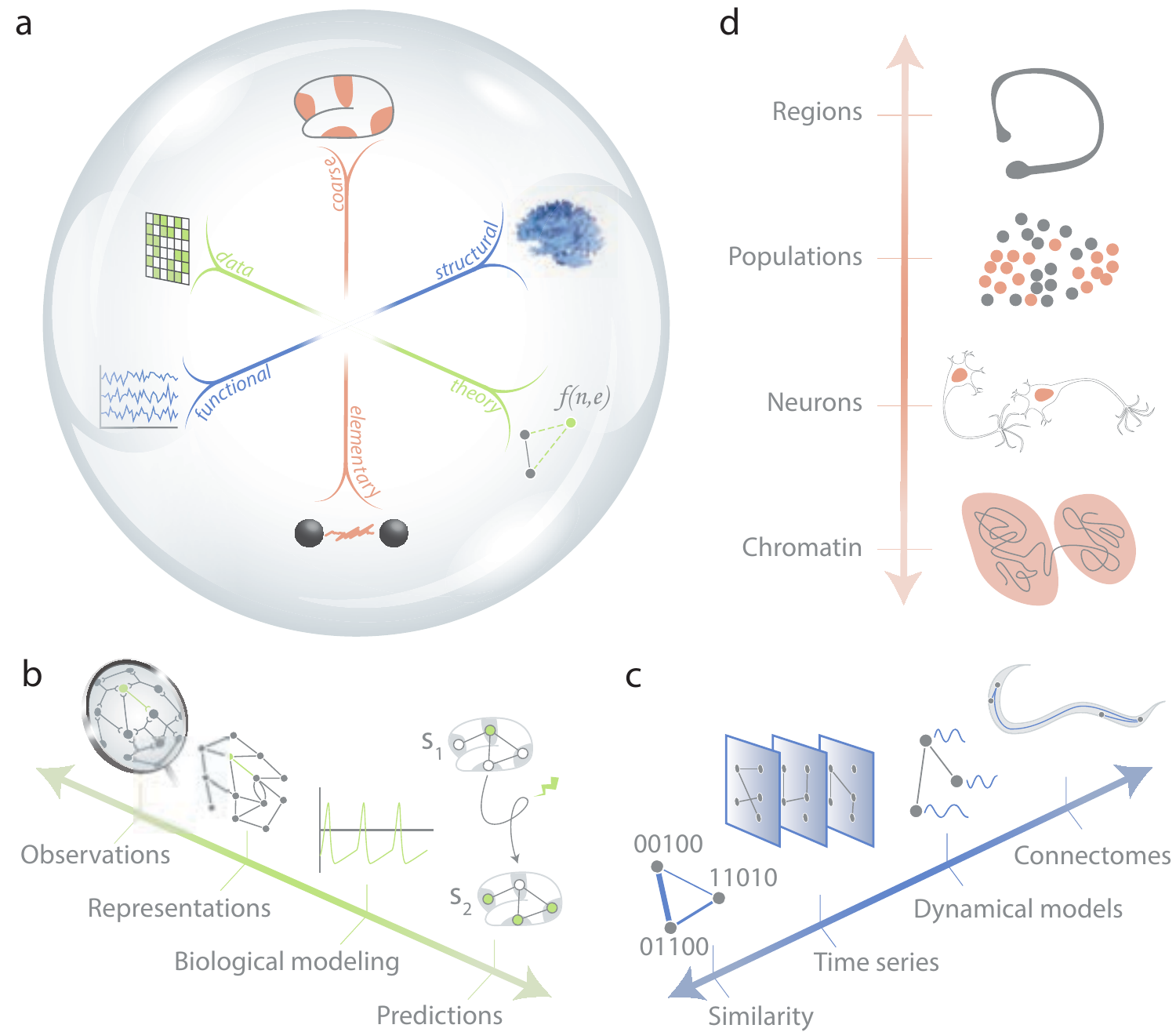}
\caption{\textbf{Three dimensions of network model types.} \emph{(a)} We posit that efforts to understand mechanisms of brain structure, function, development, and evolution in network neuroscience can be organized along three key dimensions of model types. \emph{(b)} The first dimension extends from elementary descriptions to coarse-grained approximations. \emph{(c)} The second dimension extends from biophysical realism to functional phenomenology. \emph{(d)} The third dimension extends from data representation to first-principles theory. Figure adapted with permission from \cite{bassett2018nature}.}
\label{fig2}
\end{figure*}

Network neuroscience as a field is devoted to building, exercising, and validating network models of neural systems with the explicit goal of better understanding brain structure and function, as well as cognition, behavior, and disease \cite{medaglia2015cognitive,sporns2014contributions,stam2014modern,braun2018from,fornito2017opportunities}. The types of network models that are built share a similar analogical basis that emphasizes the importance of network-based architectures across spatial and temporal scales. However, these models differ from one another in many important ways that directly impact the sorts of inferences that can be justifiably drawn from them. Here we briefly describe recent efforts to systematize the study of network models in neuroscience by organizing these similarities and differences according to three dimensions that reflect the model categories described above \cite{bassett2018nature}: first, their phenomenological basis, ranging from representations of measured phenomena to first-principles theory; second, their target of idealization, from biophysical to functional features; and third, their scalar focus, ranging from elementary descriptions to coarse-grained approximations (Fig.~\ref{fig2}).

The dimension from data representation to first-principles theory is arguably the most fundamental to network modeling efforts in neuroscience \cite{abbott2008theoretical}. Modeling efforts of the former type begin with empirically acquired data. They then seek to build a representation of those data by stipulating which part of the data to represent as a network node, and which part of the data to represent as a network edge. Intuitively, the data representation provides an abstract, non-visual depiction or description of the system (for a few early examples, see \cite{scannell1995analysis,young1994analysis,hilgetag2000hierarchical,kaiser2006nonoptimal,stam2004functional,achard2006resilient,fallani2006brain,micheloyannis2006small,bettencourt2007functional,watts1998collective,sporns2005human}). In contrast, to make a prediction about system behavior either now or in the future, one must turn to models that instantiate first-principles theories. These models combine a network with a mathematical expression specifying the dynamics of network nodes, network edges, or collections of nodes and/or edges (see, for example, \cite{breakspear2017dynamic,melozzi2017virtual,bezgin2017mapping,ritter2013virtual,roy2014using,falcon2016new,gu2015controllability,kim2018role,yan2017network}). Data-driven network models enjoy the benefits of biological realism, whereas theory-based models have the capacity to make predictions and unearth function.

The dimension from biophysically to functionally defined features differentiates models that are physical in nature from those that are statistical in nature. Network models with biophysical realism are composed of nodes that represent physical units, including neurons, cortical columns, or Brodmann areas; and of edges that represent physical links, including synapses, projections, or white-matter tracts (see, for example \cite{varshney2011structural,nicosia2013phase,bassett2010efficient,oh2014mesoscale,kaiser2006nonoptimal,sporns2005human}). Network models addressing functional phenomenology are comprised of nodes and edges that are not necessarily physically instantiated but may instead be defined as statistical abstractions \cite{bettencourt2007functional,achard2006resilient,vandiessen2015opportunities,chu2012emergence,burns2014network,khambhati2016virtual}. Common examples of such abstract edges are those that offer estimates of effective connectivity or functional connectivity, the latter of which are also referred to as noise correlations \cite{friston2011functional,brody1999correlations,brody1999disambiguating}. It is often important to distinguish between these two types of models because they have distinct utility in assessing a network's physical constitution versus inferring its functional capacities.   

The dimension from elementary descriptions to coarse-grained approximations is critical to support a multiscale understanding of brain structure, function, and dynamics. In general, network models can encode the organization of interconnections among cells, ensembles, cortical columns or subcortical nuclei, and large-scale brain areas. As evidenced by the diversity of scales represented in current empirical and theoretical investigations, no single level of description can provide a complete explanation for cognitive function and behavior. Yet in many cases, it is worthwhile or at least practical to consider a single scale for a given study and to use insights gained at that scale to inform larger theories of multiscale function. The challenge in developing an appropriate network model at a particular scale is to ensure that network nodes represent well-defined, discrete, non-overlapping units, and that network edges represent organic, irreducible relations \cite{butts2009revisiting}. Whereas models at the final spatial scale consider elementary building blocks \cite{feldt2011dissecting,betzel2017diversity,kim2015fast,sautois2007role,teller2014emergence,tang2008maximum,kaiser2017mechanisms,mahadevan2017living}, models at the coarse spatial scale consider emergent functions \cite{breakspear2017dynamic}.

\begin{figure}[ht]
\centering
\includegraphics[width=0.9\linewidth]{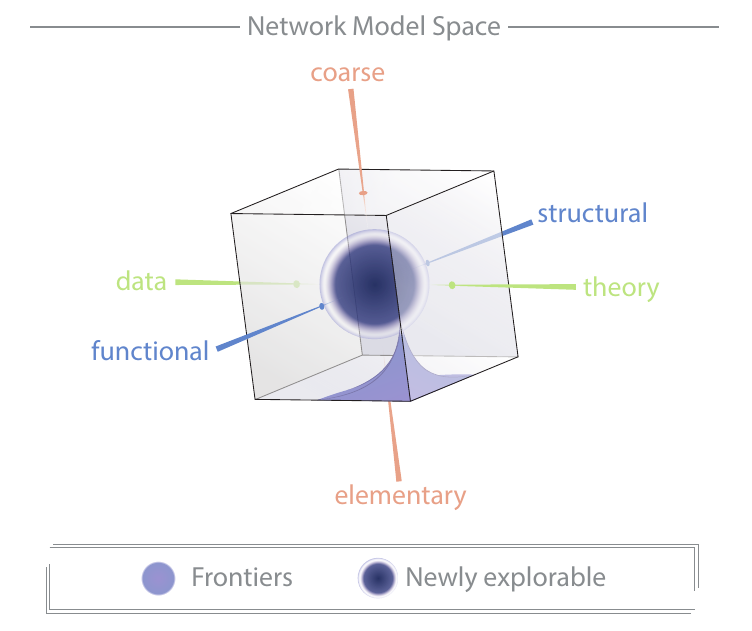}
\caption{\textbf{Density of study in the three dimensional space of network models.} Density plot showing varying levels of study across the 3-dimensional space of model types in network neuroscience. The center of the 3-dimensional space (aubergine) is becoming increasingly accessible due to empirical and computational advances over the past few decades. The least well-studied space (periwinkle) represents first-principles theories of functional phenomenology at the elementary level of description. Clear volumes indicate spaces that are most commonly studied. }
\label{fig3}
\end{figure}

Together, these complementary dimensions define a three-dimensional space of network models that can be used to enhance our understanding of brain structure and function (Fig.~\ref{fig3}). Notably, prior modeling efforts have not been pursued with equal vigor in all volumes of this space, partly reflecting historical factors and the changing state of neurotechnologies. Early work focused on large-scale network models of anatomy, which represent coarse-grained data representations with biophysical realism  \cite{bassett2006small,bassett2016small,liao2017small}. Less well-studied are first-principles theories of functional phenomenology, particularly among elementary units. With a marked increase in the pace of data acquisition and the capacities for data analysis, we anticipate increasing success developing network models at the center of the volume. In other words, we anticipate a wider array of studies that inform first-principles theories with data, complement physical models with statistical inferences on informational capacity, and build explicit multiscale accounts of network function. These multi-faceted network models will enhance our ability to explain different parts, processes, or principles of the nervous system.

\section*{Modeling perturbations to networks}

When building network models, we are often concerned with demonstrating their validity \cite{bassett2018nature}. In prior work, we followed canonical principles for validating animal models of disease to suggest that network models can display three distinct types of validity \cite{schmueli2010explain,willner1984validity,belzung2011criteria,mckinney1969animal}: descriptive, explanatory, and predictive. Intuitively, descriptive validity requires that the model resemble the system under study, a concept akin to face validity in animal models \cite{willner2017reliability}. For example, a model with descriptive validity might accurately reflect the specific pattern of nodes and edges observed in anatomical or functional data. By contrast, explanatory validity requires that the model can be used to define statistical tests, for example by assessing causal relations based on the network's architecture. Finally, predictive validity is attained when there is a correlation between a network model's response to perturbation and an organism's response to that same perturbation \cite{belzung2011criteria}. Such perturbative studies can be operationalized using stimulation, lesion, ablation, or drugs. 

Predictive validity is often the final goal of any scientific domain of inquiry \cite{shneider2009four}. Because predictive validity depends on understanding its response to a perturbation, it is of interest to consider the different ways in which a network model can be perturbed. Recent work in the physics and engineering communities has begun to focus on how the architecture of a network determines how perturbations affect its function. A simple way in which to parse these studies is to consider separately perturbations applied: (i) to a single node or to a single edge (``point perturbations''), (ii) to a set of nodes or to a set of edges, and (iii) across a fixed area or volume of the network's topology. In the context of network neuroscience, these different types of perturbations may be accessible to distinct empirical techniques and collectively could be used to better understand both endogenous and exogenous control, thereby informing clinical intervention \cite{tang2018control}.   

From a modeling perspective, point perturbations are perhaps the simplest type to study. Initial modeling approaches focused point perturbations in the form of node or edge removal. At times referred to as virtual lesioning, this approach was developed to quantify the robustness of a network by estimating the difference between the value of a graph statistic estimated before node or edge removal, and the value of that same graph statistic estimated after node or edge removal (Fig.~\ref{fig1}b) \cite{dong2013robustness}. When nodes are removed at random, the approach is referred to as a \emph{random attack}. When nodes are removed based on their topological role in the network, estimated using the values of various graph statistics such as degree and betweenness centrality, the approach is referred to as a \emph{targeted attack} \cite{achard2006resilient}. These approaches have recently been used to better understand the impact of regional dysfunction in schizophrenia and stroke \cite{lynall2010functional,lo2015randomization,alstott2009modeling}.

Other approaches address how a perturbation of the activity of a node or edge can change the activity of other parts of the network. This approach is central to network control theory and built on the foundations of linear systems theory \cite{motter2015networkcontrology,kailath1980linear}. Here one considers the pattern of interconnections between units as well as a model of dynamics that specifies how the activity at one node can travel along edges to other nodes in the graph \cite{tang2018control}. By modeling both the connectivity and the dynamics, one can identify ``driver'' nodes with time-dependent control that can guide the system's activity \cite{liu2011controllability}.
A recent application of these techniques to the connectome of \emph{C. elegans} demonstrated that a particular network model had striking validity in predicting the effects of single cell ablasions on the organism \cite{yan2017network}. The approach can also be extended beyond the identification of drivers controlling all dynamics to the identification of drivers controlling specific dynamics \cite{pasqualetti2014controllability}. The specificity of this extension allows for the study of unique control strategies within neural systems. A recent application of this technique provided an explanation for the anatomical location of areas of the brain involved in executive function, as those most capable of enacting modal controllability \cite{gu2015controllability}.   

Despite their analytical tractability, point perturbations can be the most difficult to enact and interpret in the context of real neural systems. On a conceptual level, a single, functional node or edge used in a model might not have an obvious, well-defined anatomical substrate in the brain to target. On a practical level, even given a well-defined target, it may not be possible to cleanly perturb just that target given the lack of complete specificity associated with current microstimulation, optogenetic, and pharmacological methods. Nonetheless, point perturbations represent a useful starting point in considering the validation of network models.

Moving beyond point perturbations, it is also of interest to consider perturbation to multiple points in the network, or to entire areas or volumes of neural systems. Intuitively, multi-point control is a natural reflection of circuit activity, where several areas may be activated simultaneously to orchestrate a change in communication or dynamics \cite{palmigiano2017flexible}. Multi-point control could also be fruitfully applied to the development of stimulation therapies to quiet seizure dynamics using implantable devices \cite{ridder2017state,ehrens2015closed,taylor2015optimal,jobst2017brain}. Fortunately, the general network control framework is readily extended to account for the activity of multiple control points simultaneously, and can be used to directly model the propagation of stimulation along white matter tracts to predict distant effects on regional activity \cite{muldoon2016stimulation,stiso2018white}. Extending these tools to affect control over continuous areas or volumes of a network is more difficult, and remains an important area for future work. Progress in this area is critical for extending network models to account for other chemical mechanisms of transcellular communication and the effects of glia, neuromodulatory systems, and other mechanisms on brain function and behavior \cite{borroto2015role,saytchouk2018gliotransmission,safaai2015modeling,bruinsma2018relationship}.

\begin{figure*}[ht]
\centering
\includegraphics[width=0.75\linewidth]{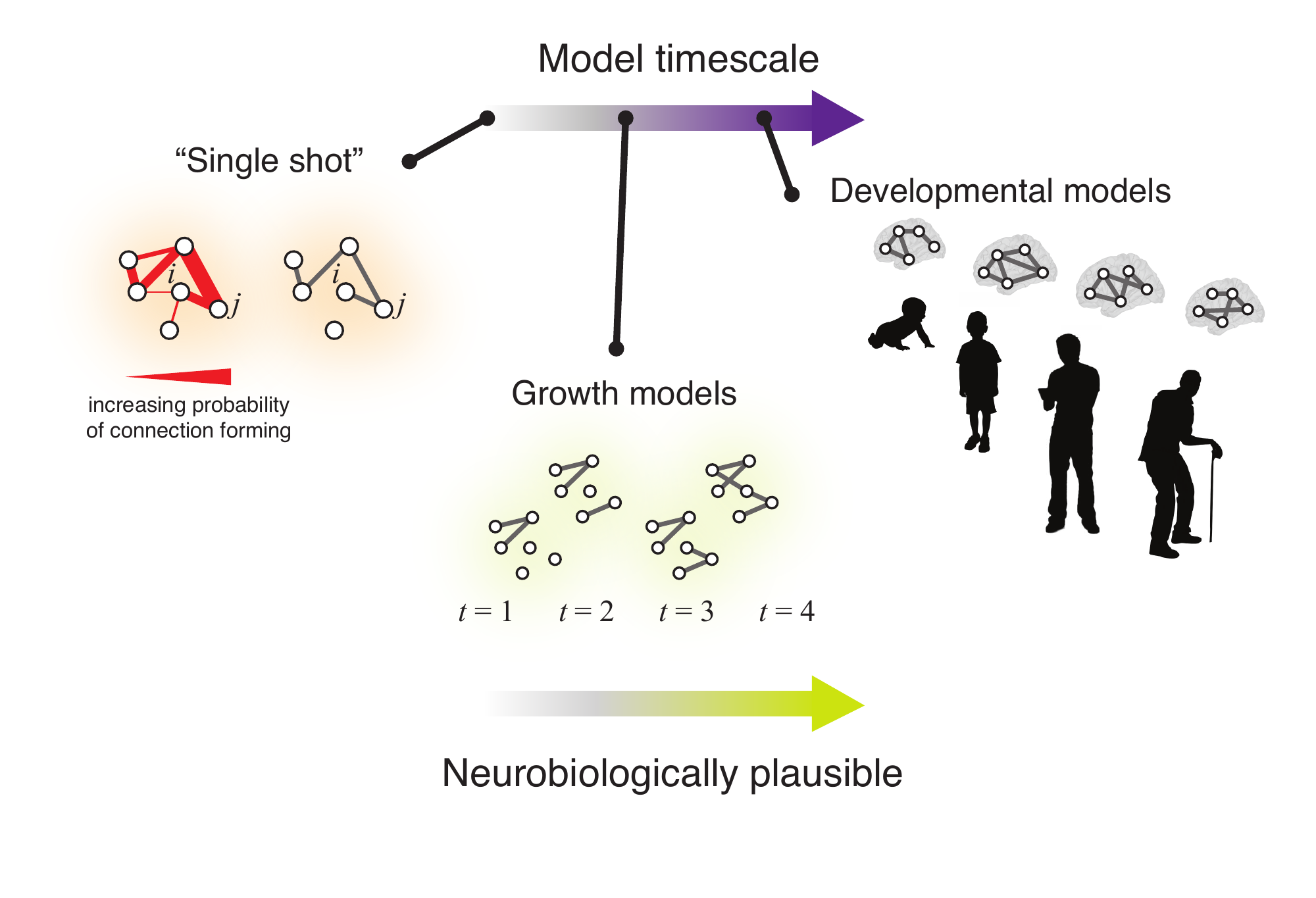}
\caption{\textbf{Distinct classes of generative network models.} Generative network models exist in three main classes that differ in the timescales over which they operate. \emph{(Left)} Single-shot models specify a functional form for the probability with which any two nodes are linked with one another. \emph{(Middle)} Growth models specify rules by which nodes and/or edges are added to the network over time, which is commonly discretized in arbitrary units. \emph{(Right)} Developmental models specify wiring rules with fixed timescales in actual units of seconds, minutes, days, months, years, etc. in an effort to better match the true growth mechanisms of an organism \cite{nicosia2013phase}. Figure reproduced with permission from \cite{betzel2017generative}.}
\label{fig4}
\end{figure*}

\section*{Modeling network growth and evolution}

The study of network perturbations, while useful for understanding endogenous and exogenous mechanisms of control, is also pertinent to an understanding of how neural systems came to be, how they develop, and how they age. A change in gene expression can alter the natural progression of cell fate from pluripotent stem cell through neuroprogenitor cell and eventually neuron \cite{mahadevan2017living}. A fluctuation in chemical gradients can comprise a perturbation that alters the course of neuronal migration, and by extension the location and density of synapses \cite{wrobel2014directed}. In fully developed adult neurons, Hebb's rule essentially postulates that perturbations to neuronal firing can alter cellular-level network architecture \cite{bi2001synaptic}. Even at the large scale in humans, long-term training can induce changes in white matter architecture evident in non-invasive neuroimaging \cite{scholz2009training}. Understanding how perturbations of the organism or part of the organism over both short and long time scales affect network growth and evolution is an important open area of research.

Some progress has been made over the last few years in constructing so-called generative network models. Such models stipulate a wiring rule in the hope of producing a network architecture that displays topological or functional properties that are similar to those observed in the networks representing real systems \cite{betzel2017generative}. The basic idea is that a wiring rule that produces a network topology similar to that observed in the real system is a candidate mechanism for network generation. The inference is made stronger if the wiring rule also displays characteristics thought to be consistent with biology, such as parsimony and efficiency. A common way of testing the pragmatic utility of the generative network model is to determine if it can be used to make out-of-sample predictions about held-out network data. 

Generative network models tend to be built in one of three types: single-shot models, growth models, and developmental models (Fig.~\ref{fig4}). A single-shot model specifies a form for the connection probabilities, from which all edges and their weights are then drawn \cite{vertes2012simple,vertes2014generative,betzel2016generative,beul2015predictive,beul2017predictive,hilgetag2016primate}. A growth model specifies a time-dependent wiring rule that indicates how nodes and possibly even edges are added over time \cite{klimm2014resolving}. Developmental models extend the biological realism of the effort even farther by specifying wiring rules in which the time scales of the model match the time scales of development in the organism under study \cite{nicosia2013phase}. Together, these three types of generative network models vary in the time scale over which they operate and in their neurobiological plausibility.

Recently, generative network models have been developed and applied to explain neurophysiological and neuroanatomical data across both elementary descriptions and coarse-grained approximations \cite{beul2015predictive,betzel2016generative}. For example, a particularly striking single-shot model of the neural connectome of the nematode \emph{C. elegans} demonstrated that a wiring rule based on the random outgrowth of axons, in combination with a competition for available space at the target neuron, was able to recapitulate the empirical network's edge length distribution \cite{kaiser2009simple}. A developmental model of the same organism combined information regarding the pattern of interconnectivity between neurons with information regarding the birth times of neurons and their spatial locations \cite{nicosia2013phase}. This study provided compelling evidence for a trade-off between the network's topology and cost that appears to be differentially negotiated over different developmental time periods. With a few exceptions \cite{vertes2012simple}, most generative network models have focused on data representations more so than first-principles theories, and biophysical realism more so than functional phenomenology. Expanding efforts to fill the full space of model types will be an important area for future work in generative modeling.

\section*{Future directions}

In considering the future utility of network models for advancing our understanding of neural systems, it is worth pointing out that the models used to date are relatively simple from a mathematical perspective. It remains an open question whether more complex network models might prove useful or merely obfuscate inference. To address this question, one could rationally assess whether some aspect of a known neurophysiological process remains unaccounted for by existing models. For example, to increase descriptive validity, one might wish to build an annotated network, where nodes can be assigned values or properties, reflecting for example cerebral glucose metabolism estimates from fluorodeoxyglucose (FDG)-positron emission tomography (PET), blood-oxygen-level dependent (BOLD) contrast imaging, magnetoencephalographic (MEG) or electroencephalographic (EEG) power, gray matter volume or cortical thickness, or cytoarchitectonic properties \cite{newman2016structure,murphy2016explicitly}. Furthermore, to increase explanatory validity, one might wish to build multilayer networks where the nodes and edges in each layer are obtained from different types of measurements \cite{kivela2014multilayer,twarie2014structural,yu2016building}, and where the architecture of the network is allowed to vary over time in concert with system function \cite{holme2012temporal, khambhati2017modeling,sizemore2017dynamic,kopell2014beyond,breakspear2017dynamic}. Such richer models could allow one to test how network dynamics at one time point or in one modality might cause a change in network dynamics at another time point or in another modality.

A second way in which to address the question of whether more complex network models might prove useful is to consider whether the testing of a particular hypothesis requires a novel network model. For example, recent efforts have provided initial evidence that some higher-order, non-pairwise interactions occur between neurons and between large-scale brain areas \cite{ganmor2011sparse,lord2016insights}. Critically, all of the network models we have discussed here are based on pairwise interactions and cannot directly account for non-pairwise interactions \cite{sizemore2018importance,petri2014homological}. Tools that have been developed in the applied mathematics community that can encode and characterize higher-order relations include hypergraphs (an edge can link any number of vertices) and simplicial complexes (higher-order interaction terms become fundamental units) \cite{bassett2014cross,giusti2016twos}. These generalizations of graphs may be critical for an accurate understanding of neuronal codes and associated computations both at micro- and macro-scales \cite{sizemore2017cliques,reimann2017cliques,curto2013neural}. As the field moves beyond univariate accounts to postulate more network-based hypotheses, richer network models may be required.

\section*{Conclusion}
From cellular to regional scales, neural circuitry is an interconnected system. In such a system, network modeling is a particularly useful approach for distilling interconnection patterns into tractable mathematical objects that are amenable to theory. Here we discussed the nature of network models, which share a similar networked architecture that is justified in terms of its analogies to brain structure but then differ along several dimensions: from data representations to first-principles theory, abstractions that emphasize biophysical or functional features, and different scales from elementary descriptions to coarse-grained approximations. We paid particular attention to models that have been developed to better understand the response of networked systems to perturbation enacted at a single-point, at multiple points, or across extended areas or volumes of the organism. We also offered an extended discussion of generative network models that seek to identify candidate wiring mechanisms for circuit evolution or development. We suggest that network models are particularly appropriate for neural systems. Accordingly, future advances in our understanding of computation and cognition will depend upon the expansion of these models in mathematical sophistication, and the development of richer, network-based hypothesis of brain structure, function, and dynamics.  

\section*{Acknowledgements} 
We thank Blevmore Labs and Ann E. Sizemore for efforts in graphic design. We also thank D. Lydon-Staley, A. E. Sizemore, E. Cornblath and D. Zhou for helpful comments on an earlier version of this manuscript. D.S.B. acknowledges support from the John D. and Catherine T. MacArthur Foundation, the Alfred P. Sloan Foundation, the ISI Foundation, the Paul Allen Foundation, the Army Research Laboratory (W911NF-10-2-0022), the Army Research Office (Bassett-W911NF-14-1-0679, Grafton-W911NF-16-1-0474, DCIST-W911NF-17-2-0181), the Office of Naval Research, the National Institute of Mental Health (2-R01-DC-009209-11, R01-MH112847, R01-MH107235, R21-M MH-106799), the National Institute of Child Health and Human Development (1R01HD086888-01), National Institute of Neurological Disorders and Stroke (R01 NS099348), and the National Science Foundation (BCS-1441502, BCS-1430087, NSF PHY-1554488 and BCS-1631550). J.I.G. acknowledges support from the National Science Foundation (NSF-NCS 1533623), the National Eye Institute (R01-EY015260), and the National Institute of Mental Health (R01-MH115557). The content is solely the responsibility of the authors and does not necessarily represent the truth.

\section*{Competing Interests}
The authors declare no competing interests. 

\newpage
\bibliography{bibfile}

\end{document}